# High Density, Localized Quantum Emitters in Strained 2D Semiconductors

Gwangwoo Kim[a], Hyong Min Kim[a], Pawan Kumar[a,b], Mahfujur Rahaman[a], Christopher E. Stevens[c,d], Jonghyuk Jeon[e], Kiyoung Jo[a], Kwan-Ho Kim[a], Nicholas Trainor[f], Haoyue Zhu[f], Byeong-Hyeok Sohn[e], Eric A. Stach[b], Joshua R. Hendrickson[c], Nicholas R Glavin[g], Joonki Suh[h], Joan M Redwing[f], Deep Jariwala[a,*]

[a]*Department of Electrical and Systems Engineering, University of Pennsylvania, Philadelphia, PA 19104, USA.*

[b]*Department of Materials Science and Engineering, University of Pennsylvania, Philadelphia, PA 19104, USA.*

[c]*Air Force Research Laboratory, Sensors Directorate, Wright-Patterson Air Force Base, OH 45433, USA.*

[d]*KBR Inc., Beavercreek, OH 45431, USA.*

[e]*Department of Chemistry, Seoul National University, Seoul 08826, Republic of Korea.*

[f]*2D Crystal Consortium-Materials Innovation Platform, Materials Research Institute, The Pennsylvania State University, University Park, PA 16802, USA.*

[g]*Air Force Research Laboratory, Materials and Manufacturing Directorate, Wright-Patterson Air Force Base, OH 45433, USA.*

[h]*Department of Materials Science and Engineering, Ulsan National Institute of Science and Technology, Ulsan 44919, Republic of Korea.*

KEYWORDS: Transition metal dichalcogenides, Tungsten diselenide, Strain engineering, Platinum nanoparticles, Quantum emitter

**\*Corresponding Authors**

*E-mail Addresses:* E-mail: dmj@seas.upenn.edu

**ABSTRACT**

Two-dimensional chalcogenide semiconductors have recently emerged as a host material for quantum emitters of single photons. While several reports on defect and strain-induced single photon emission from 2D chalcogenides exist, a bottom-up, lithography-free approach to produce a high density of emitters remains elusive. Further, the physical properties of quantum emission in the case of strained 2D semiconductors are far from being understood. Here, we demonstrate a bottom-up, scalable, and lithography-free approach to creating large areas of localized emitters with high density (~150 emitters/um$^2$) in a WSe$_2$ monolayer. We induce strain inside the WSe$_2$ monolayer with high spatial density by conformally placing the WSe$_2$ monolayer over a uniform array of Pt nanoparticles with a size of 10 nm. Cryogenic, time-resolved, and gate-tunable luminescence measurements combined with near-field luminescence spectroscopy suggest the formation of localized states in strained regions that emit single photons with a high spatial density. Our approach of using a metal nanoparticle array to generate a high density of strained quantum emitters opens a new path towards scalable, tunable, and versatile quantum light sources.

**INTRODUCTION**

Strain engineering and physical confinement transformation engineering of three-dimensional (3D) semiconductors has been studied either for bandgap engineering or for the confinement of discrete quantum states for quantum information processing[1-3]. Importantly, the utilization of local strain for generating quantum confinement has been achieved extensively in III–V semiconducting quantum dots[4, 5], nanowires[2], and nanotubes[6]. Recently, such studies have been extended to two-dimensional (2D) materials, primarily in insulating hexagonal boron nitride[7-10] and semiconducting transition metal dichalcogenides (TMDCs)[11-14]. 2D materials provide a promising platform for strain engineering due to their unique structure. Their strong in-plane bonding makes them amenable to hetero-integration with arbitrary substrates that contain other optical/photonic nanostructures. This implies that strain can be readily tuned via the substrate topography. To achieve this, Si-based nanopillars[15-18] have been widely used as 3D templates (substrates) to induce such localized strains. Moreover, as these mechanical strains can strongly control band structure, it is possible to use mechanical strain to tune electronic and photonic performance[19, 20]. For this reason, the identification of quantum emitters (QEs) in TMDCs has generated considerable excitement in the field of 2D nanophotonics[21-23] and quantum information science and engineering[24]. Despite these intriguing properties, the fundamental origin of quantum emission in TMDCs is not completely clear, and thus far, it is believed that emissions are formed by excitons bound to defects[12-14], impurities[11], or 3D transformations (nanostructures, nanoindents) induced by the strain gradients[15-18, 25-28]. In addition to this fundamental question, there is a need to create a high spatial and number density of quantum emitters in 2D semiconductors. The high number density and high brightness of individual quantum emitters are need to create dense integration of quantum light sources with other photonic elements in scalable applications.

Here, we demonstrate an approach to forming strain-induced localized quantum emitters with high spatial density (> 150 /um$^2$) in a 2D TMDC monolayer that is placed on the top of an array of uniform metal nanoparticles (NPs) with AlO$_x$ dielectric layers using a top-down lithography free approach. We observe localized exciton (LX) emission from the Pt NP-strained WSe$_2$ structures by using far-field photoluminescence (PL) spectroscopy at room temperature. In addition, we perform a systematic study that combines low-temperature far-field and room temperature near-field PL spectroscopy to arrive at a mechanistic understanding of the LX emission from strained WSe$_2$. We conclude that the LX emission originates from the radiative emission control of dark excitons by compressive strain. We also demonstrate an identical LX emission in a large area-WSe$_2$ monolayer grown by metal organic chemical deposition (MOCVD), which is spatially controllable by using a lithographically patterned array of Pt NPs. Cryogenic and time-resolved PL measurements suggest lifetimes of ~11 ns and with < 0.7 nm spectral linewidths of emission, which further confirms that our samples have a high density of single photon emitters formed by a top-down, lithography-free process.

## RESULTS AND DISCUSSION

**Fabrication of local-strained WSe$_2$ monolayer and its characterization.** The WSe$_2$ monolayers are transferred over a uniform array of Pt NPs deposited with a thin AlO$_x$ spacer. The sample configuration and a schematic of the fabrication process are illustrated in **Figure 1a** and **Figure S1**, respectively. The thin AlO$_x$ layer was first deposited on an array of Pt NPs spread over a SiO$_2$/Si substrate, prepared with the aid of self-patterning diblock copolymer micelles[29]. The Pt NPs were synthesized by spin-coating a single layer of polystyrene-block-poly(4-vinylpyridine) (PS-P4VP) micelles with hexachloroplatinic acid (H$_2$PtCl$_6$) precursor for Pt NPs within their cores, followed by annealing at 400 °C. Depending on the molecular weight of the diblock copolymer, the size of the Pt NPs was further controlled to two different sizes with average diameters of 11.7 and 5.9 nm (See **S2 and S3a-b** for additional characterization). Most of the results discussed below were obtained on an 11.7 nm-sized Pt NPs array with an average inter-particle distance of 103 nm. Upon synthesis of the array of Pt NPs, we deposited a thin layer of AlO$_x$ (2, 5 nm thickness, **Figure S3c-f**) for use as a spacer via atomic layer deposition (ALD). This spacer layer is necessary to prevent quenching of excitons in 2D TMDCs via direct contact with metallic Pt NPs. It is worth noting that thicker AlO$_x$ depositions result in larger height differences between the top of Pt NPs and the substrate, suggesting preferential ALD growth of AlO$_x$ on the Pt NPs. Following ALD AlO$_x$ deposition, either mechanically exfoliated (ME) or MOCVD-grown WSe$_2$ monolayers were transferred onto the AlO$_x$-deposited Pt NP arrays by either a pick-up (for exfoliated crystals) or a wet-transfer (for MOCVD grown films) method. Finally, the as-prepared samples were annealed at 300 °C with Ar flow (50 sccm) in a vacuum tube furnace to make better conformal contact by removing trapped solvent or gas molecules at the interface between the WSe$_2$ and AlO$_x$/Pt NPs substrate. Structural characterization of locally-strained WSe$_2$ on AlO$_x$/Pt NP arrays after annealing was performed by AFM (**Figure 1b,c**) and scanning transmission electron microscopy (STEM) (**Figure 1d,e** and **Figure S4,5**), respectively. As shown in **Figure 1b**, we clearly observe visible wrinkles between NPs, indicating deformation and strain in the WSe$_2$ monolayers. Atomically resolved STEM image (**Figure 1e**) and energy dispersive X-ray spectroscopic (EDS) elemental mapping (**Figure S5**) show that the WSe$_2$ monolayer is conformal to the AlO$_x$-covered Pt NP array, which is direct evidence that a locally-strained structure has been created.

The effect of local strain on the prepared WSe$_2$ monolayer is examined by micro-PL spectroscopy with a 405 nm excitation laser. **Figure 1f** compares the representative PL spectra of the locally-strained WSe$_2$ (red) and unstrained WSe$_2$ (black) layers, measured with a 100x lens (0.9 numerical aperture, NA) at room temperature. As reported in literature precedent[17, 30, 31], the strained WSe$_2$ exhibits a clear redshift (~30 meV) of the neutral exciton (X$^0$) peak (1.66 eV). From the magnitude of the redshift (X$^0$) and the theoretical calculation[32], we estimate our strained WSe$_2$ samples are subjected to approximately 0.38 ± 0.021% biaxial tensile strain. The presence of strain in the WSe$_2$ was confirmed by micro-Raman spectroscopy (**Figure S6**). The strained WSe$_2$ sample (**Figure S6a**) shows a significant redshift of 0.65 cm$^{-1}$ for the E$_{2g}$/A$_{1g}$ Raman peak, again indicating ~0.35 ± 0.034% biaxial tensile strain[17], in good agreement with the strain values derived from PL peak shifts above. We further confirmed the strain effect via Raman spectroscopy in another TMDC: WS$_2$ (**Figure S6b**). In that case, the in-plane E$_{2g}$ mode also shows a similar strain-dependent redshift, whereas the out-of-plane A$_{1g}$ mode showed negligible shift under applied strain. Note that these tensile strains are concentrated on the top of Pt NPs and compressive strain may be formed between the NPs. This will be discussed in more detail below. Interestingly, we observe a new LX emission in our strained samples, in addition to the X$^0$ emission (Figure 1f red curve). The energies of the observed LX emission (1.4-1.55 eV) are lower than that of X$^0$ (1.63 eV), similar to the recently reported results for WSe$_2$ nanobubbles[33]. **Figure 1g** presents the temperature-dependent PL spectra of the strained WSe$_2$ monolayer on the AlO$_x$(5 nm)/Pt NP array. Note that the LX peak, as seen in **Figure 1f** red curve at room temperature, was not visible in the room temperature spectrum (red) in **Figure 1g** because of the instrumental configuration, which utilized a long distance-50x lens (0.35 NA) for variable temperature measurements in a vacuum and cryogenic stage. However, we still clearly observed the LX peaks in the 1.6-1.65 eV range at low temperature (80 K), while no LX emission was observed in the unstrained WSe$_2$ on bare SiO$_2$/Si substrate (**Figure S7**) across the entire temperature range. Similarly, the LX emission was also confirmed in a MOCVD-grown WSe$_2$ monolayer formed over the entire area, shown in **Figure S8**.

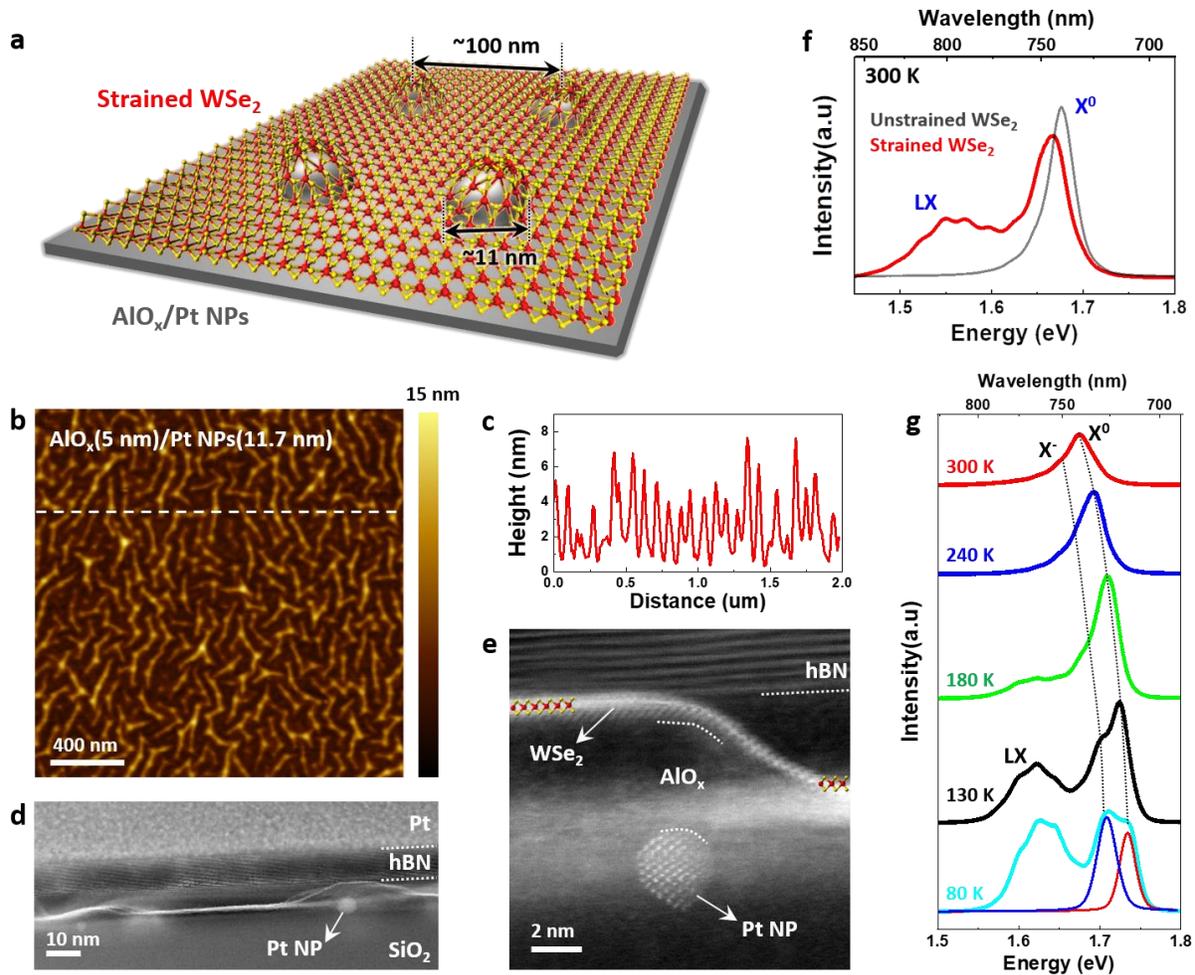

**Figure 1.** Structural characterization and observation of localized emission in strained WSe$_2$ monolayers. (a) Schematic depiction of strained WSe$_2$ monolayer on AlO$_x$/Pt NP array with the relevant dimensions labelled. (b) AFM height image of strained WSe$_2$ monolayer on AlO$_x$(5 nm-thick)/Pt NPs(11.6 nm-size). (c) Height profile marked with the white line in Figure b. (d-e) Representative cross-sectional STEM images of the strained WSe$_2$ monolayer with h-BN protecting layers clearly showing the curved topography that induces local strain as well as the buried Pt NP with visible atomic columns. (f) Room-temperature PL spectra of strained (red) and unstrained (black) WSe$_2$, placed on AlO$_x$(5 nm)/Pt NPs/SiO$_2$ and bare SiO$_2$ substrate, respectively. These spectra were obtained by 405 nm-laser excitation and a 100x lens with 0.9 NA. (g) Temperature-dependent PL spectra of strained WSe$_2$ on AlO$_x$(5 nm)/Pt NPs/SiO$_2$. These spectra were obtained by 405 nm-laser excitation and a 50x lens with 0.35 NA.

**Spatial- and magnitude-controllable localized emission in strained WSe$_2$ monolayer.** The above observations, particularly in MOCVD-grown WSe$_2$, suggest that it is possible to pattern regions showing LX emissions over a large area. Since the strain and the resulting LX emission from the TMDC are highly dependent on the substrate topography, it can be spatially controlled by patterning the Pt NP array on the substrate. To demonstrate this, we prepared stripe-patterned arrays of Pt NPs on a substrate using photolithography to mask and selectively etch Pt NPs with aqua regia solution (**Figure S9**) in the exposed regions. Then, after depositing the AlO$_x$ thin layer by ALD, a MOCVD-grown WSe$_2$ monolayer was transferred to the patterned substrate and annealed in the same condition as discussed above. **Figure 2a** shows AFM height images of the alternating strained/unstrained MOCVD-WSe$_2$ films. As shown in the magnified image (**Figure 2b**), the presence and absence of a Pt NP array is clearly identified. To verify the origin of the LX emission, we performed far-field PL analysis at 80 K temperature. Notably, the PL mapping image (**Figure 2c**) in the range of 750−825 nm clearly confirms that the LX emission is only detected in the strained WSe$_2$ area. The red and blue lines in **Figure 2d** show the corresponding PL spectra of the strained and unstrained regions, respectively. These results provide clear evidence that the origin of the LX peak in emission is related to the strain effect on the WSe$_2$ monolayer.

To understand the strain effect and its relation to the LX emission, we performed low temperature-PL measurements on several control samples. First, we note that although there is a distinct possibility that defects or inducing impurities are formed in the WSe$_2$ during the annealing treatment. However, no LX emission was observed in the unstrained WSe$_2$ sample on a flat SiO$_2$/Si substrate (**Figure S7**), which was annealed and went through the same treatments as the strained samples. Second, it is also plausible that the origin of the LX peak might be due to either charge-trapped sites in the AlO$_x$ dielectric layer or hybridization of localized surface plasmons in the Pt NPs with excitons in WSe$_2$. To verify that these effects were not contributing in this system, we prepared two kinds of control WSe$_2$ samples: on (i) AlO$_x$-deposited SiO$_2$ substrate without Pt NPs (**Figure S10**) and (ii) pristine Pt NP array/SiO$_2$ substrate (**Figure S11**). The LX emission was not observed in either sample: The sample (i) at low temperature showed no visible LX emission, while in sample (ii), the PL emission completely disappeared due to direct contact between WSe$_2$ and metallic Pt after annealing, which resulted in quenching of the emission. These control experiments further confirm that the LX emission is likely due to the local strain effect in the WSe$_2$ monolayers.

Given that our strain-inducing agent is an array of Pt NPs, this results in the ability to control these local strains by varying the size and spacing of Pt NPs and the deposition thickness of the AlO$_x$ layer. Inducing these changes in the substrate topography allows altering the intensity change of LX emission. **Figure 2d** shows the strained WSe$_2$ monolayer on AlO$_x$ (5 nm-thick)/Pt NPs with smaller size (5.9 nm) and spacing (51 nm) (**Figure 2b**). As shown in the AFM height images, we observed that the surface was relatively non-uniform and congested compared to the previous AlO$_x$(5 nm)/Pt NPs (11.7 nm) sample(**Figure 1b**). Smaller and denser Pt NP arrays created stronger and denser strain, leading to an enhanced LX emission. (**Figure 2h**). Next, as mentioned previously (**Figure S3**), the height of the Pt NPs could be changed by controlling the AlO$_x$ thickness during the deposition process, which is expected to affect the strain in the WSe$_2$ monolayer placed on top. **Figure 2f** shows an AFM image of a strained WSe$_2$ monolayer on AlO$_x$ (2 nm)Pt NPs (11.7 nm). This sample had a relatively smooth surface and, as a result, LX emission with a smaller intensity was observed at 80 K. Also, the LX peaks tended to shift to slightly lower energy with increasing strain. This will be discussed in more detail later.

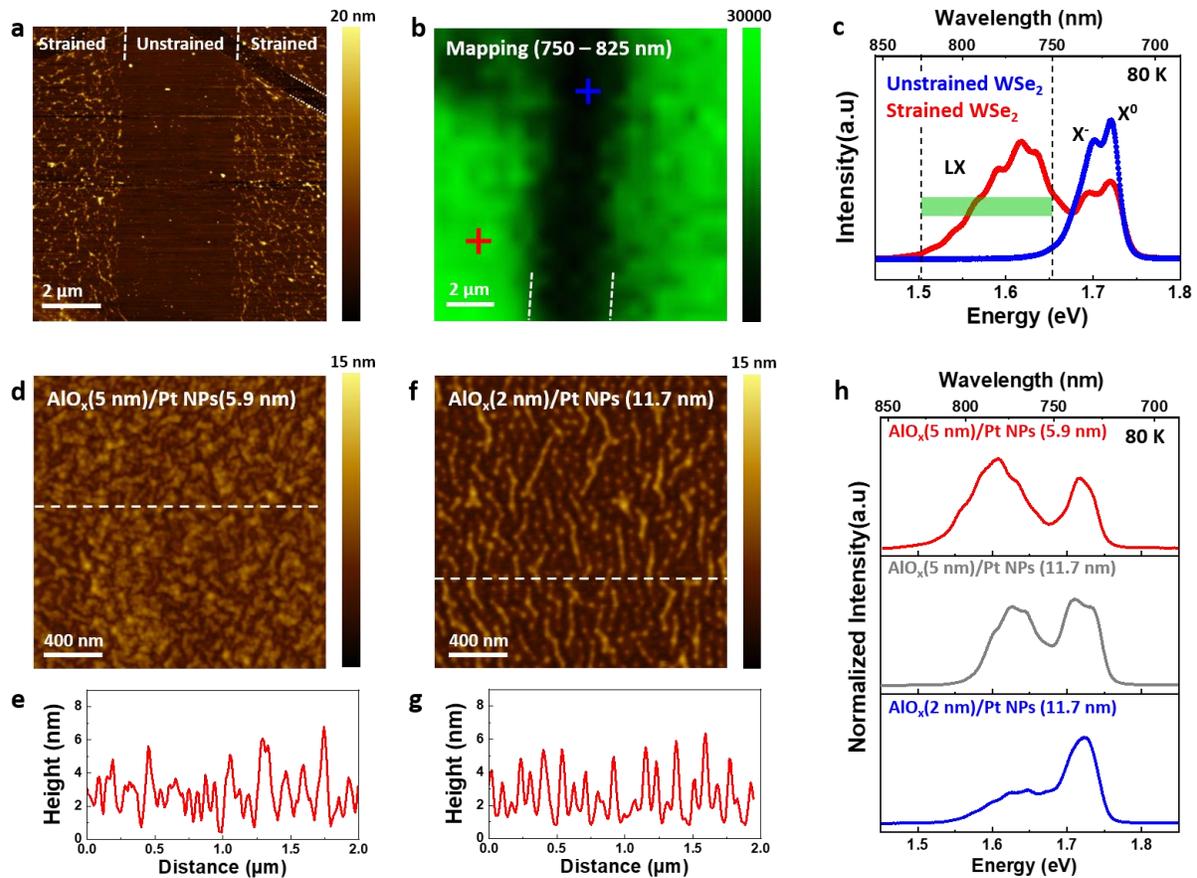

**Figure 2.** Spatial- and magnitude-controllable localized emission in a strained WSe$_2$ monolayer by a patterned Pt NP array. (a) AFM height image of MOCVD-grown WSe$_2$ monolayer placed on the stripe-patterned Pt NP array with 5 nm-thick-AlO$_x$. (b) Far field-PL mapping of localized emission energy (integrated intensity from 750 – 825 nm) at 80 K temperature. (c) PL spectra from the corresponding points marked with blue (unstrained WSe$_2$) and red (strained WSe$_2$) crosses in Figure b. (d,f) AFM height images of WSe$_2$ monolayers on (d) AlO$_x$(5 nm)/small Pt NPs(5.9 nm)/SiO$_2$ and (f) AlO$_x$(2 nm)/large Pt NPs(11.7 nm)/SiO$_2$. (e,g) Height profiles marked with the white line in the Figure d and f. (h) Low-temperature PL spectra of WSe$_2$ monolayers on each substrate.

**Nano-optical imaging and spectroscopic analysis of localized emissions in strained WSe$_2$.**
The far-field PL spectroscopic results between room temperature and 80 K described above show that the observed LX emission is due to strain in the WSe$_2$ monolayer induced by the Pt NP array substrate. To verify the spatial position of LX emission and mechanistically understand the origin of the LX emission, we performed tip-enhanced PL spectroscopy (TEPL) on our samples with a 633 nm-excitation laser and a spatial resolution of ~ 10 nm. **Figure 3** displays TEPL results on the strained WSe$_2$ monolayer on AlO$_x$(2 nm)/Pt NPs/SiO$_2$ substrate. Note that TEPL measurement was carried out at room temperature and a relatively thin AlO$_x$ layer (2 nm) was deposited to allow strong confinement of the gap-plasmon mode between the tip and the underlying Pt NPs such that the PL signal from the WSe$_2$ monolayer sandwiched between the Au tip and Pt NPs is enhanced by virtue of the gap plasmonic cavity. As shown in the AFM height image (**Figure 3a**), WSe$_2$ flexure is distributed with a size of < 25 nm on the surface. **Figure 3d** contrasts the TEPL spectra collected from region of the WSe$_2$ monolayer center above a Pt NP (blue) from the periphery (red). The blue spectrum of the Pt NPs center is composed of the typical PL emission of a WSe$_2$ monolayer with a neutral exciton ($X^0$) peak. The red spectrum measured around the Pt NPs is noticeably different, exhibiting the LX emission, similar to the previously shown far-field PL spectrum (red, **Figure 1f**). Interestingly, the corresponding TEPL map (**Figure 3c**) in the range of 775-850 nm shows that the LX emission is localized to a concentric ring around the Pt NPs rather than on the top of the NP, where the tensile strain is largest. This implies that compressive strains (yellow arrows) formed on the curved WSe$_2$ monolayer (i) between the top of Pt NPs or (ii) in contact with the flat substrate may be the likely cause of the LX emission (**Figure 3e**). These compressive strain regions are also visible in the cross-sectional TEM images (**Figure 3f,g**) and each image clearly shows the two cases (i: **Figure 3f**, ii: **Figure 3g**) mentioned above. Furthermore, the strain can be confirmed by comparing the aforementioned AFM height profiles of AlO$_x$-deposited Pt NPs (**Figure S3f**) and WSe$_2$-transferred on the top (**Figure 1c**). The difference between the highest and lowest heights decreased from 11.19 ± 1.57 nm to 6.07 ± 1.02 nm after transferring WSe$_2$. This indicates that the WSe$_2$ monolayer is freestanding, and in a bent state between the Pt NPs under compressive strain.

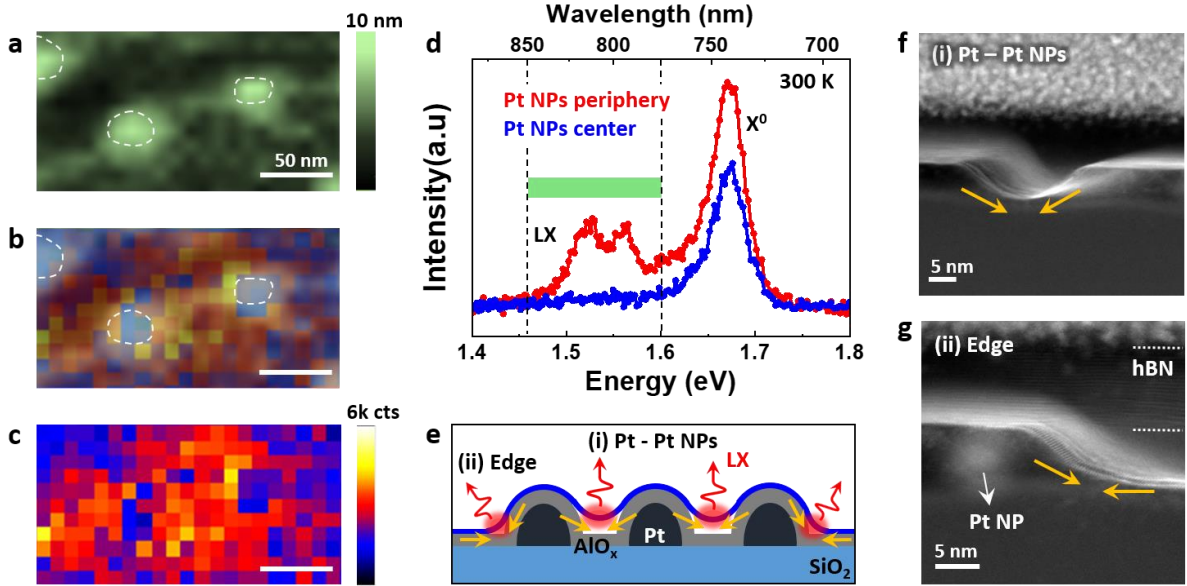

**Figure 3.** Nanoscale optical and structural imaging of localized emission on strained WSe$_2$. (a-c) AFM height image (a) and corresponding TEPL spatial map (c) of strained WSe$_2$ monolayer on AlO$_x$(2 nm)/Pt NPs/SiO$_2$. The TEPL image was created within the spectral range of 770-850 nm with a step size 10 nm. The excitation energy was 1.96 eV (633 nm). (b) Overlaid image of Figure a and c. (d) TEPL spectra of WSe$_2$ monolayer on Pt NPs center (blue) and periphery (red) as highlighted in the TEPL map. The clear differences in emission spectra at the LX peaks (1.45-1.6 eV energy range) is evident (e) Schematic diagram of the sample cross-section showing the position of localized emissions. (f,g) Cross-sectional STEM images of strained WSe$_2$ showing the two cases of the curved WSe$_2$ monolayers under compressive strain: (i, f) between the top of Pt NPs or (ii, g) in contact with the flat substrate.

To get a deeper insight into the origin of the features, we have also compared the presence or absence of the LX emission in other TMDCs and only observed the LX emissions in W-based

TMDCs and not in Mo-based TMDCs (See **Figure S12 for additional details**). From these results, we could speculate that the origin of the LX emission might be associated with the dark exciton in TMDCs. It is well known that in Mo and W-based TMDC layers, the strong spin-orbit coupling can induce the band splitting of both the conduction and valence bands into two sub-bands with opposite spins, thereby producing dark exciton states[34-36]. In the case of W-based TMDCs, these dark excitons can be easily detected in optical measurements since the dark exciton state is located energetically below the bright exciton state and is therefore the ground state of the exciton. In contrast, in Mo-based TMDCs, the bright exciton forms the ground state. Further, the bands in W-based TMDCs have a relatively large band splitting (40-50 meV), in comparison with Mo-based TMDCs (~10 meV)[37-39]. It is noteworthy that the dark-bright band splitting in our strained W-based TMDCs has a relatively large value of ~80 meV. According to a recent theoretical paper[40], the aforementioned compressive strain in TMDC monolayers enables not only to increase the dark-bright state separation value, but also to enhance the intensity of the dark exciton. This trend is clearly visible in the magnitude-controlled LX results introduced earlier (**Figure 2h**). When Pt NPs with smaller size and narrower spacing were used, or when the $AlO_x$ layer is thickly deposited, a larger compressive strain could be induced in the $WSe_2$ monolayer, making the LX peak stronger and shifted to lower energies. In addition, we also perform excitation power-dependent PL measurements to determine that the cause of the LX emission is dark excitons (See **Figure S13** for additional details). Since the intensity of the dark excitons should obey a linear dependence as a function of incident power, the power dependence is an excellent indicator for dark excitons[39, 41].

In addition to near-field PL, we also performed far-field cryogenic PL (down to 4.2 K) and time-resolved PL (TRPL) measurements to directly probe the LX emission dynamics. **Figure 4a,b** shows the PL spectra on the strained $WSe_2$ monolayer, which was obtained with a pulsed excitation laser (637 nm, 1 MHz) with 25 and 0.2 µW power at 4.2 K. As seen in the PL spectrum acquired under a high-power laser (**Figure 4a**), both $X^0$ (bright excitons) and LX emissions (dark excitons) are now clearly visible. The energy difference between $X^0$ and LX is measured to be 59.7 meV, which is higher than the previously reported values[37-39, 42]. As mentioned above, relatively large splitting originates from the radiative emission control of the dark excitons by compressive strain[40]. Also, we could observe the sharp LX emission with narrow linewidths (524 ± 102 µeV) at low excitation power (**Figure 4b**) and they originate from the dark excitons trapped in the strain-induced potential as discussed above. These sharp emission lines resemble single photon emitters from defects in TMDCs, as reported in several prior studies. While the 4.2 K PL spectra reveal sharp emission peaks indicating possible single photon emitters, to obtain another confirmation, we performed TRPL analysis (**Figure 4c**) by selecting a wavelength range in the PL signal (dash lines marked in **Figure S14**) using band-pass filters. A biexponential fit to the decay data trace shows two decay components, fast ($\tau_1$) and slow ($\tau_2$). The slow component ($\tau_2$) exhibits two orders of magnitude longer lifetime than the fast component ($\tau_1$). Given the low pumping (excitation) power employed in this measurement, the LX-state lifetime close to 11 ns corresponds to the range of previously reported lifetime values (8-12 ns)[16] of single photon emission in $WSe_2$. To further confirm single photon emission, we attempted to perform a second-order photon-correlation measurement ($g^{(2)}$) on several points on the sample. However, the large spatial density of emitters (~150 /um$^2$) within the excitation spot size prevents isolation of a single emitter from providing a reliable $g^{(2)}$ measurement. Nevertheless, based on the lifetimes and the multiple narrow linewidth emission lines emanating from the broader inhomogeneous linewidth PL spectra at low temperatures, we can indeed point to a structure with a high spatial density of single photon emitters. Similar narrow linewidth emission signals were observed in the strained $WS_2$ monolayer fabricated with the same structure (**Figure S15**), indicating that the current system can be universally applied to other 2D semiconductors with a dark ground state exciton.

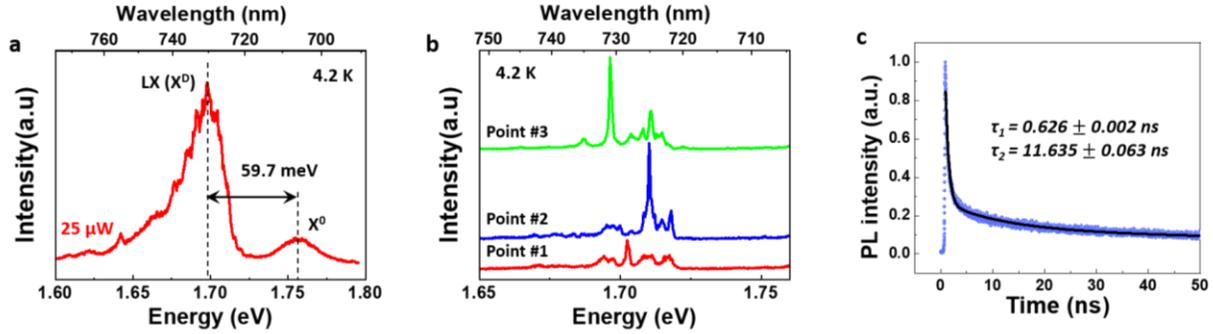

**Figure 4.** Cryogenic PL and time-resolved PL (TRPL) measurement. (a,b) PL spectra of strained WSe$_2$ monolayer with a pulse laser excitation (637 nm, 1 MHz) with an excitation power of (a) 25 and (b) 0.2 µW at 4.2 K. (c) The TRPL spectra of the localized emission. The time-resolved PL data (red dotted lines) are convoluted (black solid lines) with the instrument response function, using a biexponential function $I = A_1 \exp(-t_1/\tau_1) + A_2 \exp(-t_2/\tau_2)$.

Finally, if the strain-induced localization of exciton states is indeed true, these states should also be amenable to the electrostatic gate-induced tuning of the emission. To verify this, we fabricated a strained WSe$_2$ device encapsulated with hBN flakes through a pickup method as described previously and shown schematically in **Figure 5a**. The WSe$_2$ monolayer is gated through few-layer graphene as the top-gate electrode, with the top BN flake working as the gate dielectric (**Figure 5b**). **Figure 5c-d** shows the gate-dependent PL spectra of the strained WSe$_2$ monolayer measured at 80 K temperature. The PL spectra at specific gate voltages of +8, -8, and 0 V are shown in **Figure 5d** and the PL intensities for X$^0$, X$^+$, and LX are sensitive to the gate voltage. At the gate voltage of −8 V (red line in **Figure 5d**), the WSe$_2$ is strongly p-doped. All the intensities of X$^0$, the positive trion X$^+$, and the LX emission increase gradually, while the X$^+$ peak shows relatively stronger intensity-tunability. At the top gate voltage of +8 V (green line in **Figure 5d**), the WSe$_2$ is n-doped and all three peaks (X$^0$, X$^+$, LX) are quenched. From **Figure 5c**, **d**, it is obvious that the spectrum weight of all the resolved excitonic complexes depends sensitively on the top gate voltage that effectively controls the density and type of charge carriers in the strained WSe$_2$ monolayer.

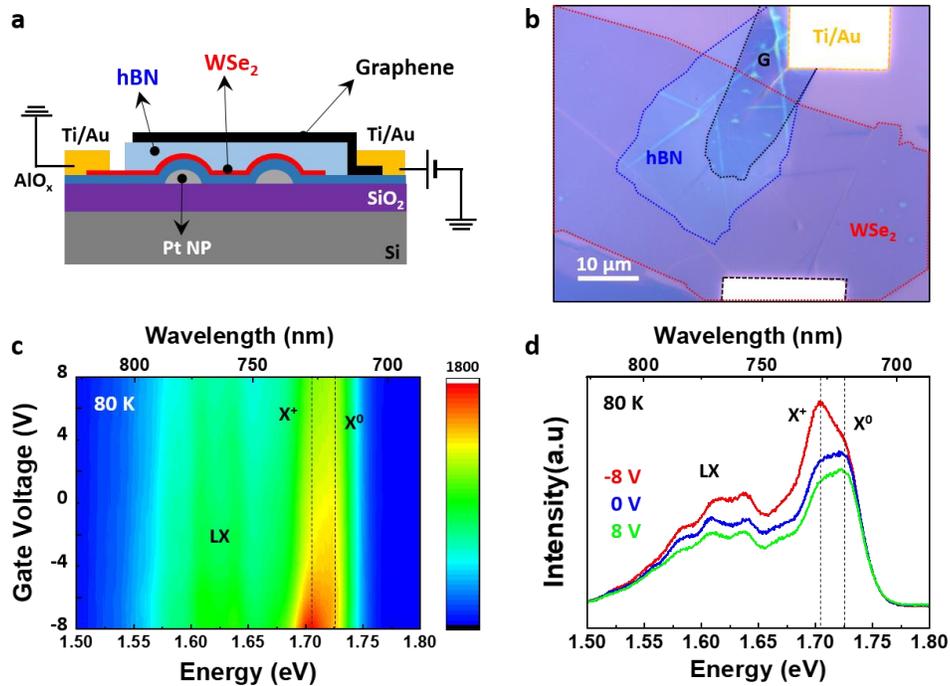

**Figure 5.** Gate-dependent PL spectroscopy of the strained WSe$_2$ device. (a,b) Schematic (a) and optical microscopic image (b) of strained WSe$_2$ device with hBN dielectric layers and graphene top gate electrode. (c) PL spectra of WSe$_2$ at 80 K as a function of the top gate voltage. The color represents the PL intensity. (d) PL spectra at the top gate voltage of +8, -8, and 0 V.

**CONCLUSIONS**

In conclusion, we have demonstrated an approach of fabricating strain-induced localized quantum emitters with high spatial density in a WSe$_2$ monolayer on a uniform Pt NP array. This strained WSe$_2$ structure shows LX at room temperature that is significantly red-shifted from both neutral excitons and trions. The sample topography and Pt NP density create compressive strain in spaces between the Pt NPs where the LX emission originates via the tuning of dark excitons by compressive strain. Our approach is free of top-down lithography and is highly scalable to large area TMDCs, as well as demonstrated to be applicable for other 2D semiconductors. Finally, our structure allows observation of quantum emission at low temperature, which is further tunable passively via Pt NP size and density as well as actively via a gate-voltage. Our results, therefore, advance the state of the art in producing a high-density of strain-induced, localized quantum emitters in TMDC monolayers in a scalable, controllable, and lithography-free approach, which is critical for future progress in the science and technology of quantum light sources.

## METHODS

**Sample preparation.** To make a uniform array of Pt NPs on a $SiO_2$ substrate, a single layer of polystyrene-block-poly(4-vinylpyridine) (PS-P4VP) micelles with an hexachloroplatinic acid ($H_2PtCl_6$) in their cores, a precursor for Pt NPs, was spin-coated on the $SiO_2$ substrate[29]. The micellar film on the $SiO_2$ was annealed at 400 °C for 30 min in air. Next, the $AlO_x$ thin layer was deposited on the Pt NP array/$SiO_2$ substrate by ALD-Cambridge Nanotech (USA) where metal organic precursor of trimethylaluminum was used with water vapor in each cycle. Then, either a mechanically exfoliated (ME) or a chemical vapor deposition (MOCVD)-grown $WSe_2$ monolayer was transferred onto the $AlO_x$-deposited Pt NP arrays by a pick-up or a wet-transfer method. ME-$WSe_2$ monolayers were exfoliated from bulk crystal (HQ-graphene) using Scotch Tape and MOCVD-grown $WSe_2$ monolayers were grown in a cold wall vertical reactor with tungsten hexacarbonyl ($W(CO)_6$, (Sigma-Aldrich, 99.99% purity)) and hydrogen selenide ($H_2Se$, Matheson, 99.998% purity)[43]. Finally, the as-prepared samples were annealed at 300 °C with Ar flow (50 sccm) in a vacuum tube furnace.

**Device fabrication.** To make a top gate device, mechanically exfoliated hBN and graphene flakes were sequentially transferred on top of the $WSe_2$ samples prepared by the above method. Next, the fabrication of Ti (10 nm)/Au (100 nm) electrode contacts was achieved by using electron beam lithography (Elionix ELS-7500EX) and an e-beam evaporator (Kurt J. Lesker PVD-75). Finally, the samples were cleaned in acetone for the lift-off process.

**Optical and structural characterization.** Far-field Raman and PL spectroscopy were performed in a Horiba LabRam HR Evolution confocal microscope with 405 nm and 633 nm-excitation lasers. The signals were collected through the 100x microscope objective (Olympus SLMPLN

N.A. = 0.90) for room temperature measurements and a 50x microscope objective (Olympus SLMPLN N.A. = 0.35) for low-temperature measurements (up to 80 K). Also, for the low-temperature analysis, samples were placed in a Linkam stage with a liquid nitrogen supply while cooling and heating and pumped to $5 \times 10^{-3}$ Torr during the measurement. Additionally, for electrostatic gating, the electrical bias was applied using a Keithley 2450 sourcemeter. An OmegaScope Smart SPM (AIST-NT) setup was used for topography scans. For near-field PL measurements, Au-coated OMNI-TERS probes (APP Nano) were used in the identical AFM setup coupled to a far-field Horiba confocal microscope with a 633 nm excitation laser. Time-resolved photoluminescence was measured by using PicoQuant's LDH-I series picosecond laser diodes (IB-470-B, IB-640-B), at a repetition rate according to the lifetime of the system. The PL was spectrally filtered using a pair of Semrock long and short pass tunable optical filters and guided into a Single Photon Avalanche Diode (Micro Photon Devices, PDM series). The event timing was recorded using a PicoQuant HydraHarp 400 event timer. The decay of the resulting timing data was then fit with an exponential (mono-, bi-, or tri-exponential, depending on which gave the best fit) using Origin's fitting software for extracting the PL lifetime. Aberration-corrected high angle annular dark field (HAADF)–STEM and EDS measurements were performed at 200 keV using JEOL NEOARM microscope. HAADF STEM images were acquired using 1 Á probe size and camera length was 4 cm. Images were captured by a Gatan annular detector using Gatan's GMS Software. EDS scan was performed utilizing large area dual detector system. Cross-sectional sample was prepared with a $Xe^+$ plasma based focused ion-beam (FIB) system (TESCAN S8000X FIB/SEM). Before cross-sectional sample preparation, mechanical exfoliated h-BN layer was dry-transferred over the sample area to soft protect the monolayer $WSe_2$ while e-beam and ion-beam deposited Pt layer was used further for hard protection to avoid any damage from the FIB processing.

AUTHOR INFORMATION

Corresponding Authors

Deep Jariwala – *Department of Electrical and Systems Engineering, University of Pennsylvania, Philadelphia, PA 19104, USA.;* E-mail: dmj@seas.upenn.edu

Author Contributions

G.K. and D.J. conceived the idea/concept. G.K. implemented the project and performed sample preparation with help of H.M.K, far-field optical (temperature dependence and gate-tunable PL up to 80 K) characterizations with help of H.M.K and K.H.K, atomic force microscopy, and data analysis. M.R. and K.J performed near-field PL analysis. P.K. performed cross-sectional sample preparation and scanning transmission electron microscopy characterization under supervision of E.A.S. and D.J. Cryogenic PL and time-resolved PL measurements at 4.2 K were recorded by C.E.S under the supervision of J.R.H. J.J. and B.H.S. provided the Pt nanoparticles array on $SiO_2$ substrates. N.T. and H.Z. prepared and characterized MOCVD-grown $WSe_2$ samples under supervision of J.M.R. G.K. and D.J. wrote the manuscript with inputs from all co-authors.

Notes

The authors declare no competing financial interest.

ACKNOWLEDGMENT

D.J. and G.K. acknowledge primary support for this work by the Air Force Office of Scientific Research (AFOSR) FA2386-20-1-4074 and partial support from FA2386-21-1-4063. D.J., E.A.S.

and P. K. acknowledge partial support from National Science Foundation (DMR-1905853) and support from University of Pennsylvania Materials Research Science and Engineering Center (MRSEC) (DMR-1720530) in addition to usage of MRSEC supported facilities. The sample fabrication, assembly and characterization were carried out at the Singh Center for Nanotechnology at the University of Pennsylvania which is supported by the National Science Foundation (NSF) National Nanotechnology Coordinated Infrastructure Program grant NNCI-1542153. K.J. was supported by Vagelos Institute of Energy Science and Technology graduate fellowship. The large area-MOCVD-$WSe_2$ monolayer samples were provided by the 2D Crystal Consortium–Materials Innovation Platform (2DCC-MIP) facility at the Pennsylvania State University which is funded by NSF under cooperative agreement DMR-1539916. J.R.H. acknowledges support from the Air Force Office of Scientific Research (Program Manager Dr. Gernot Pomrenke) under award number FA9550-20RYCOR059. M.R. acknowledges support from Deutsche Forschungsgemeinschaft (DFG, German Research Foundation) for Walter Benjamin Fellowship (award no. RA 3646/1-1). J.S. acknowledge support by the National Research Foundation of Korea (MSIT) (Grant Nos. NRF-2020R1C1C1011219 and NRF-2020M3H3A1100938). D.J. and H-M.K. acknowledge partial support from the Center of Undergraduate research Fellowships and Class of 1971 Robert J. Holtz Fund Research Grant. N.G. acknowledges the support from the Air Force Office of Scientific Research grant 19RYCOR050.

Supporting Information

High Density, Localized Quantum Emitters in Strained 2D Semiconductors

Gwangwoo Kim[a], Hyong Min Kim[a], Pawan Kumar[a,b], Mahfujur Rahaman[a], Christopher E. Stevens[c,d], Jonghyuk Jeon[e], Kiyoung Jo[a], Kwan-Ho Kim[a], Nicholas Trainor[f], Haoyue Zhu[f], Byeong-Hyeok Sohn[e], Eric A. Stach[b], Joshua R. Hendrickson[c], Nicholas R Glavin[g], Joonki Suh[h], Joan M Redwing[f], Deep Jariwala[a,*]

[a]Department of Electrical and Systems Engineering, University of Pennsylvania, Philadelphia, PA 19104, USA.

[b]Department of Materials Science and Engineering, University of Pennsylvania, Philadelphia, PA 19104, USA.

[c]Air Force Research Laboratory, Sensors Directorate, Wright-Patterson Air Force Base, OH 45433, USA.

[d]KBR Inc., Beavercreek, OH 45431, USA.

[e]Department of Chemistry, Seoul National University, Seoul 08826, Republic of Korea.

[f]2D Crystal Consortium-Materials Innovation Platform, Materials Research Institute, The Pennsylvania State University, University Park, PA 16802, USA.

[g]Air Force Research Laboratory, Materials and Manufacturing Directorate, Wright-Patterson Air Force Base, OH 45433, USA.

*h*Department of Materials Science and Engineering, Ulsan National Institute of Science and Technology, Ulsan 44919, Republic of Korea.

**\*Corresponding Authors**

*E-mail Addresses:* E-mail: dmj@seas.upenn.edu

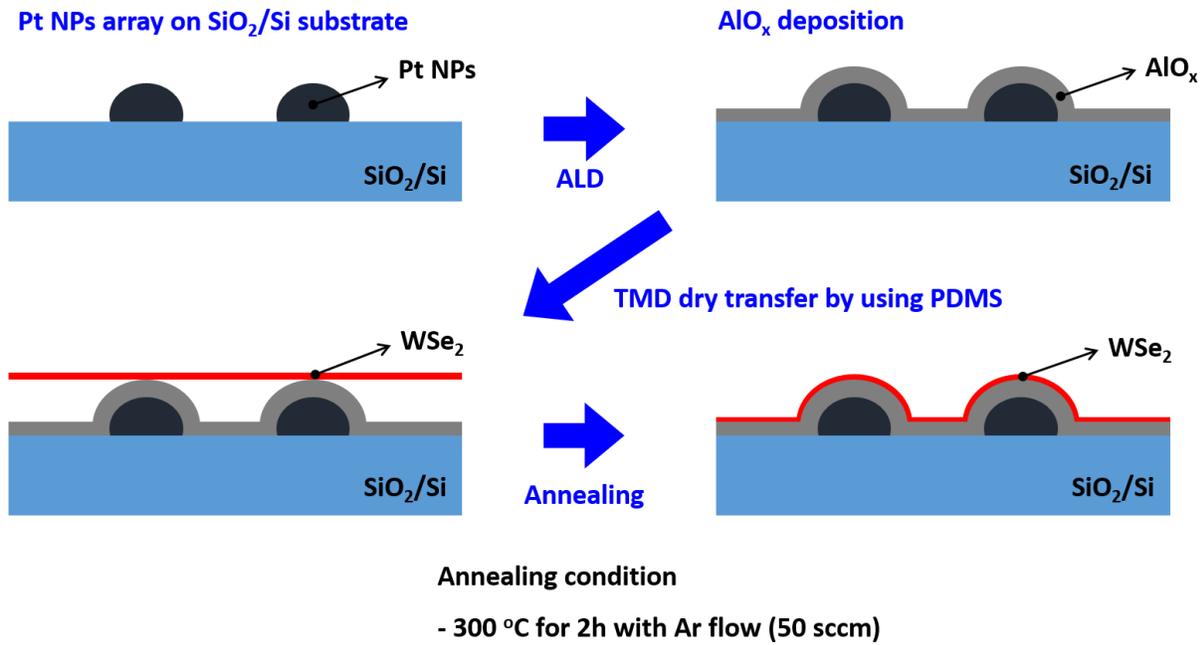

**Figure S1.** Fabrication process of local strained WSe$_2$ monolayer on AlO$_x$/Pt NPs/SiO$_2$ substrate.

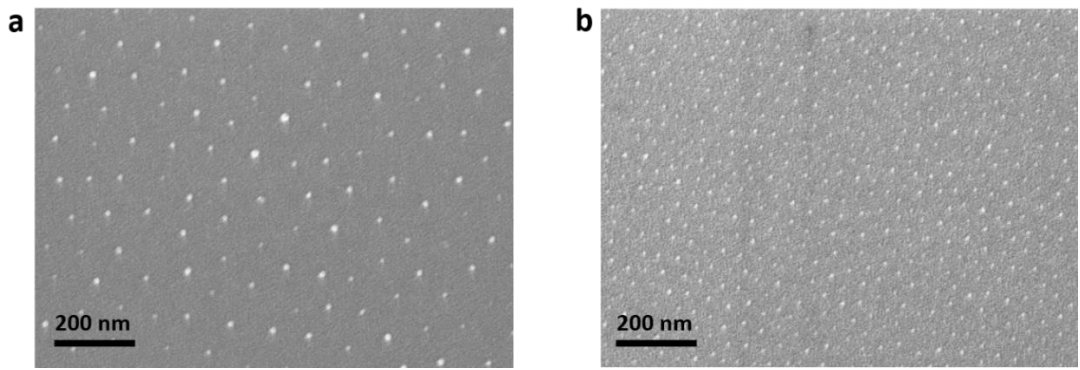

**Figure S2.** Scanning electron microscopy (SEM) image of a Pt NP array with different sizes and spacing. (a,b) SEM images of Pt NP array with different sizes and spacing. Average diameter and spacing of the larger Pt NP array (a): 11.7 nm and 103 nm. Average diameter and spacing of the smaller Pt NP array (b): 5.9 nm and 51 nm.

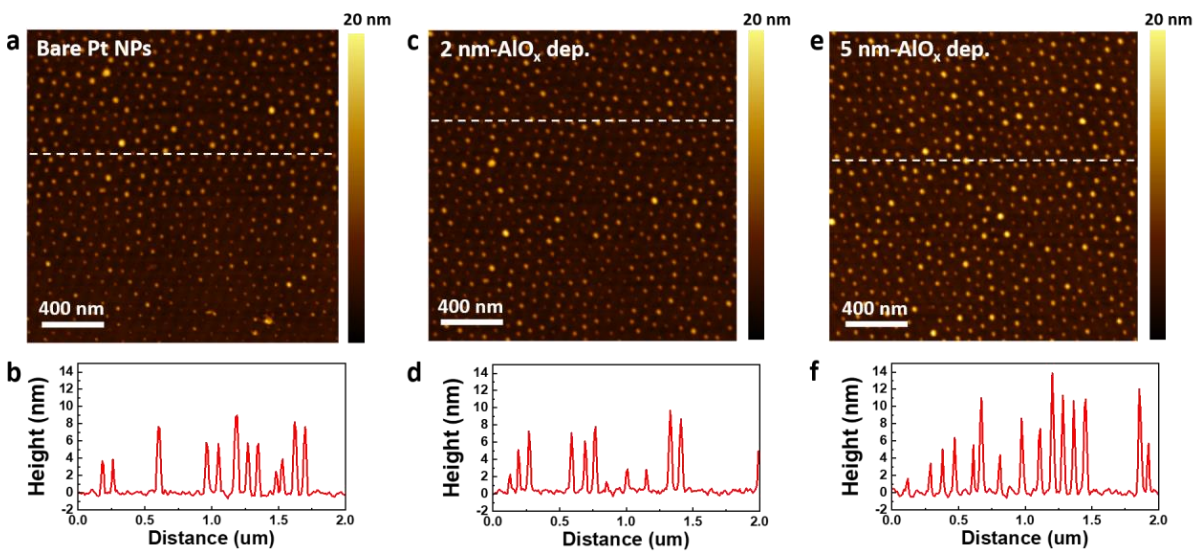

**Figure S3.** Atomic force microscopy (AFM) comparison depending on AlO$_x$ deposition thickness. (a,c,e) AFM height images of bare large Pt NPs, AlO$_x$(2 nm)/Pt NPs/SiO$_2$, and AlO$_x$(5 nm)/Pt NPs/SiO$_2$. (b,d,f) Height profiles are marked with the white line in Figures a, c, and e, respectively.

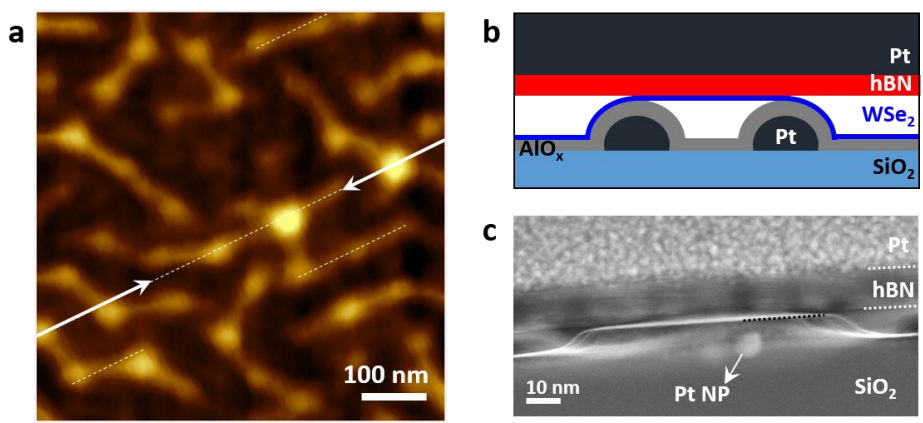

**Figure S4.** Comparison of AFM and scanning transmission microscopy (STEM) images. (a) Magnified AFM height images of Figure 1a. The white arrow indicate the perpendicular direction where focused ion-beam (FIB) processing was used to create a cross-section for STEM analysis. The white dotted lines correspond to the flat $WSe_2$ between Pt NPs, marked with a black dotted line in Figure c. (b) Depiction of the cross-section corresponding to Figure a. (c) Cross-sectional STEM image from the region shown in (a).

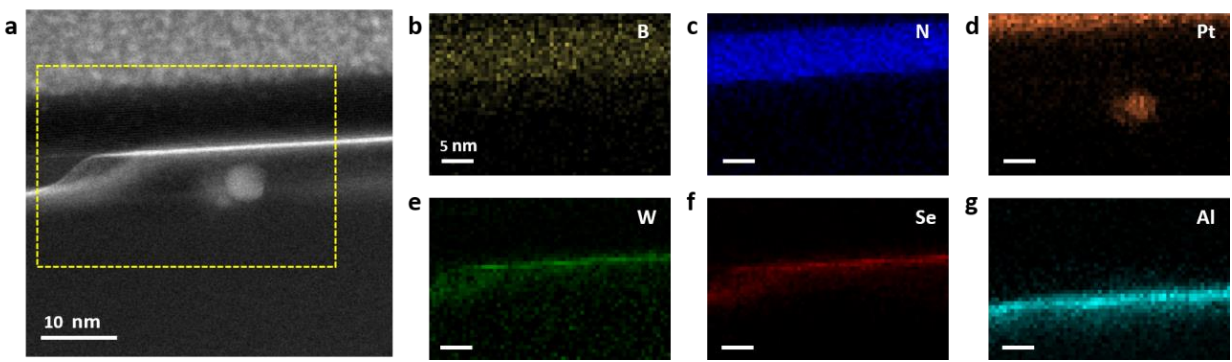

**Figure S5.** STEM EDS mapping. (a) Cross-sectional high angle annular dark field (HAADF)-STEM image of the strained $WSe_2$ on $AlO_x$/Pt NPs/$SiO_2$. (b-g) Corresponding energy dispersive X-ray spectroscopic (EDS) element mapping images of (b) boron, (c) nitrogen, (d) platinum, (e) selenium, (f) tungsten, and (g) aluminum, respectively.

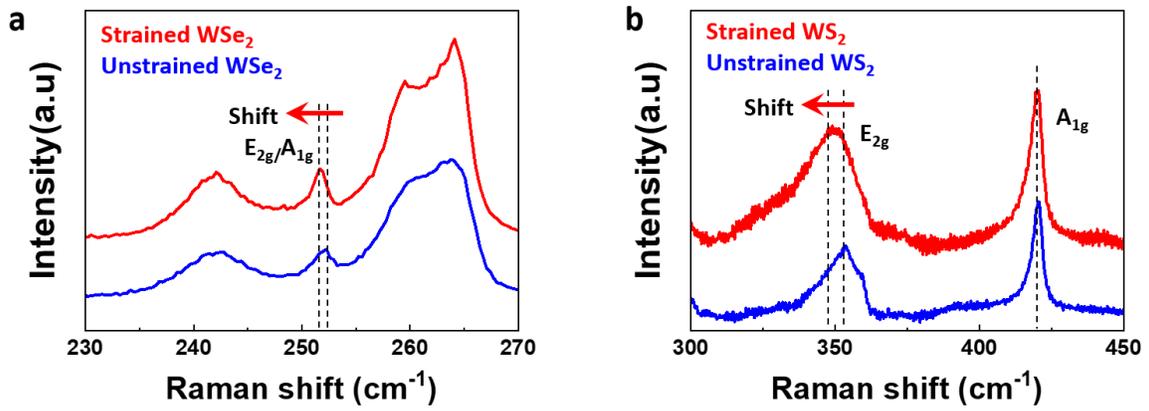

**Figure S6.** Raman comparison on strained and unstrained TMD samples. (a,b) Raman spectra of strained (red) and unstrained (blue) WSe$_2$ (a) and WS$_2$ (b) monolayer on AlO$_x$(5 nm)/Pt NPs/SiO$_2$.

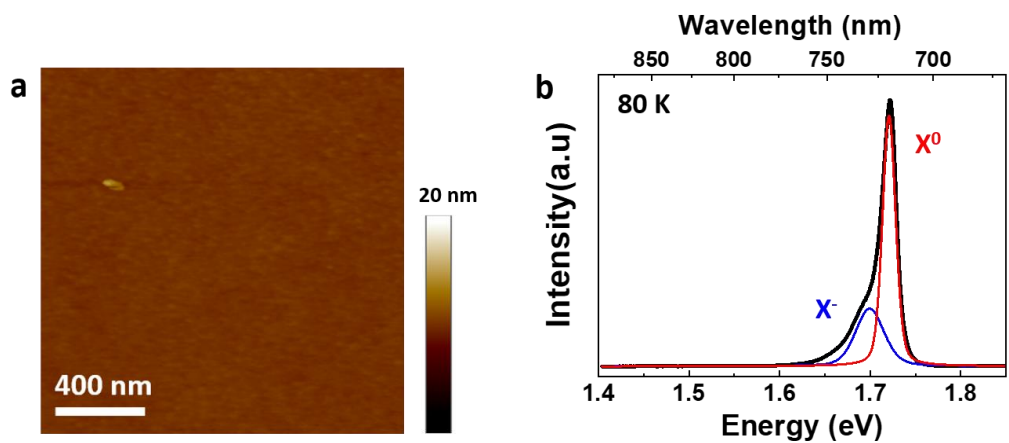

**Figure S7.** Unstrained WSe$_2$ on SiO$_2$. (a) AFM height image of WSe$_2$ on the SiO$_2$ substrate after annealing. (b) The low temperature-PL spectrum of WSe$_2$ on the SiO$_2$ substrate.

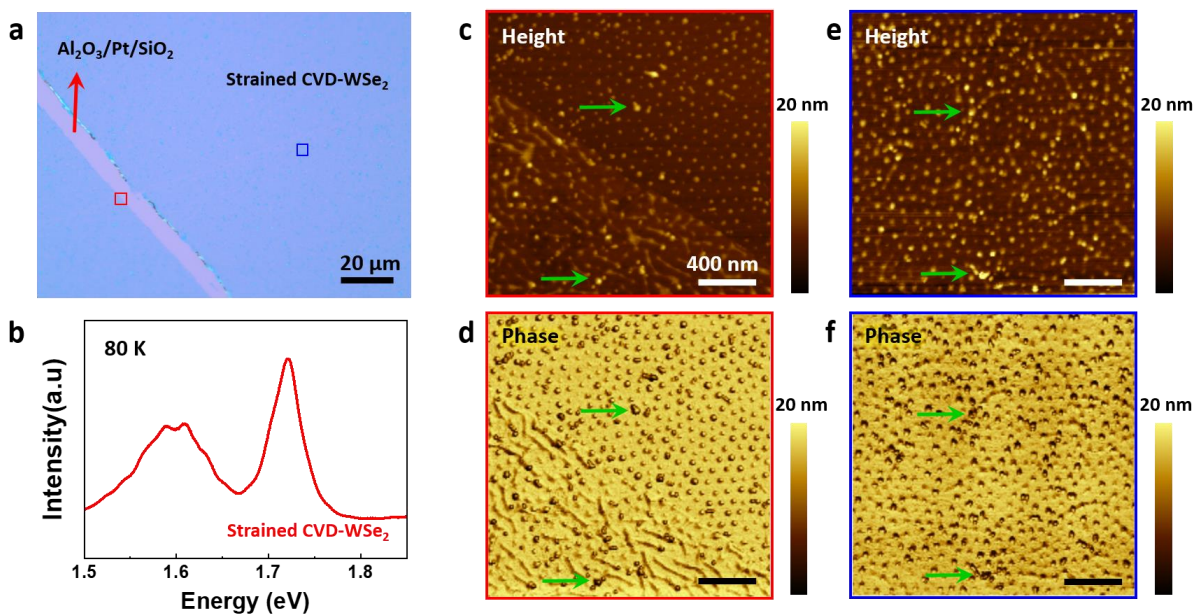

**Figure S8.** Strained CVD-grown WSe$_2$ monolayer. (a) OM image of strained CVD-grown WSe$_2$ monolayer. (b) Low temperature-PL spectrum. (c,d) AFM height (c) and phase (d) images on the edge of the WSe$_2$ monolayer. (e,f) AFM height (e) and phase (f) images on the inside area. The measurement positions are marked with red (c,d) and blue (e,f) squares in Figure a. PMMA residues are marked with green arrows.

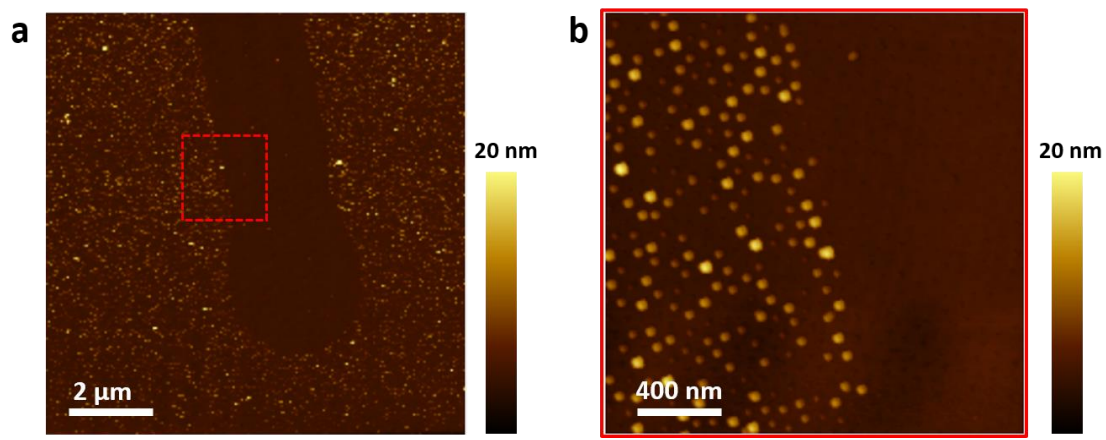

**Figure S9.** Patterned Pt NP array on SiO$_2$ substrate. (a) AFM height images of line-patterned Pt NP array on SiO$_2$ substrate by photolithography and Pt etching process (aqua regia solution). (b) The magnified image marked with the red square in Figure a.

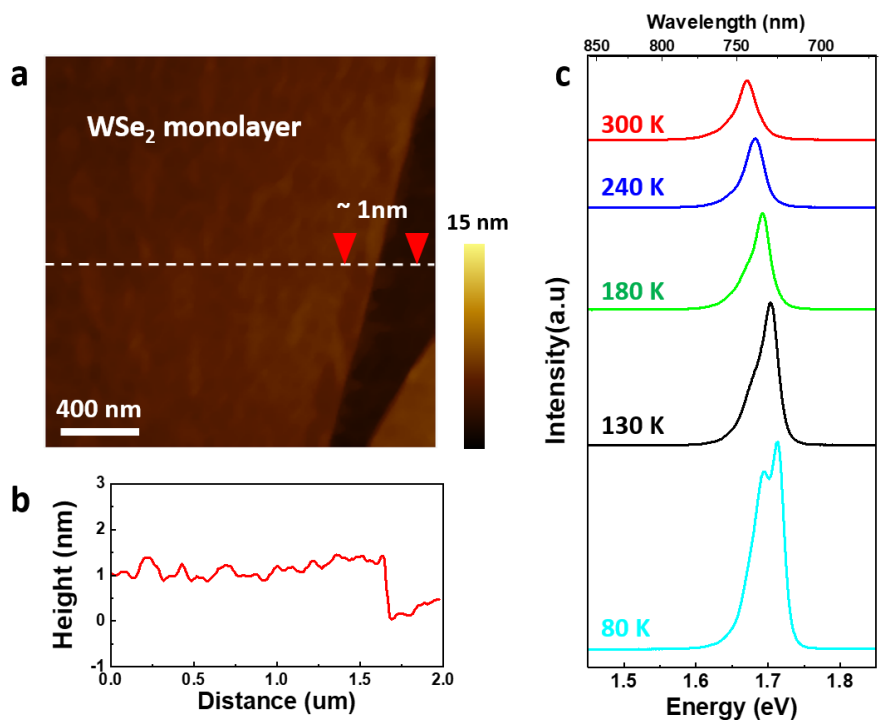

**Figure S10.** Unstrained WSe$_2$ on AlO$_x$(5 nm)/SiO$_2$. (a) AFM height image of WSe$_2$ on AlO$_x$(5 nm)/SiO$_2$ (without Pt NPs) after annealing. (b) Height profile marked with the white line in Figure a. (c) Temperature-dependent PL spectra of WSe$_2$ on AlO$_x$(5 nm)/SiO$_2$.

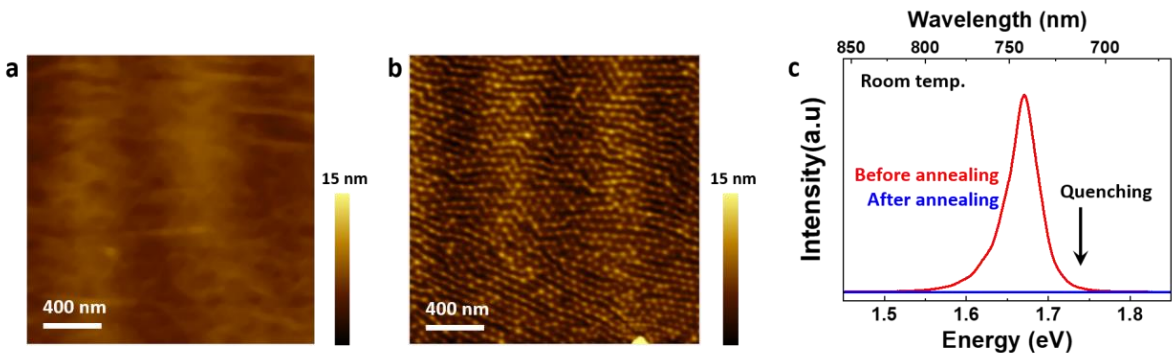

**Figure S11.** WSe$_2$ on Pt NPs/SiO$_2$. (a,b) AFM height image of WSe$_2$ on Pt NPs/SiO$_2$ (without AlO$_x$) before (a) and after (b) annealing. (c) Room temperature-PL spectra before (red) and after (blue) annealing.

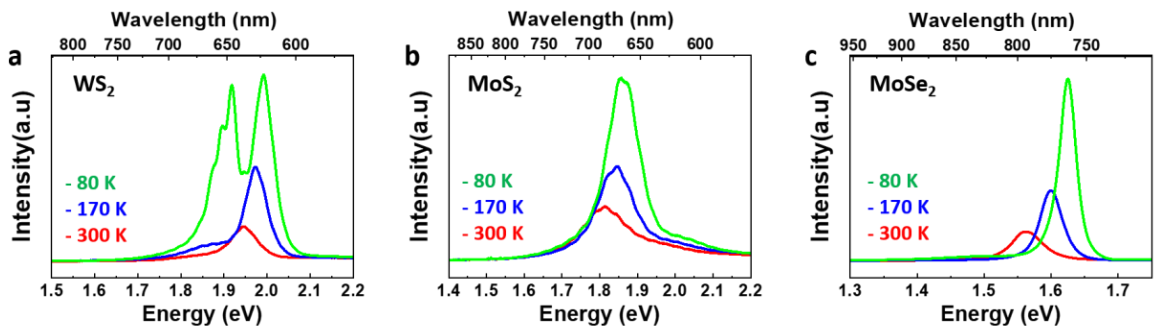

**Figure S12.** Other TMDCs (MoS$_2$, WS$_2$, MoSe$_2$) on AlO$_x$/Pt NPs/SiO$_2$ substrate. (a-c) Temperature-dependent PL spectra of strained WS$_2$ (a), MoS$_2$ (b), and MoSe$_2$ (c) monolayer on AlO$_x$(5 nm)/Pt NPs/SiO$_2$. All TMDC monolayers were prepared in the same way as the strained WSe$_2$ monolayer.

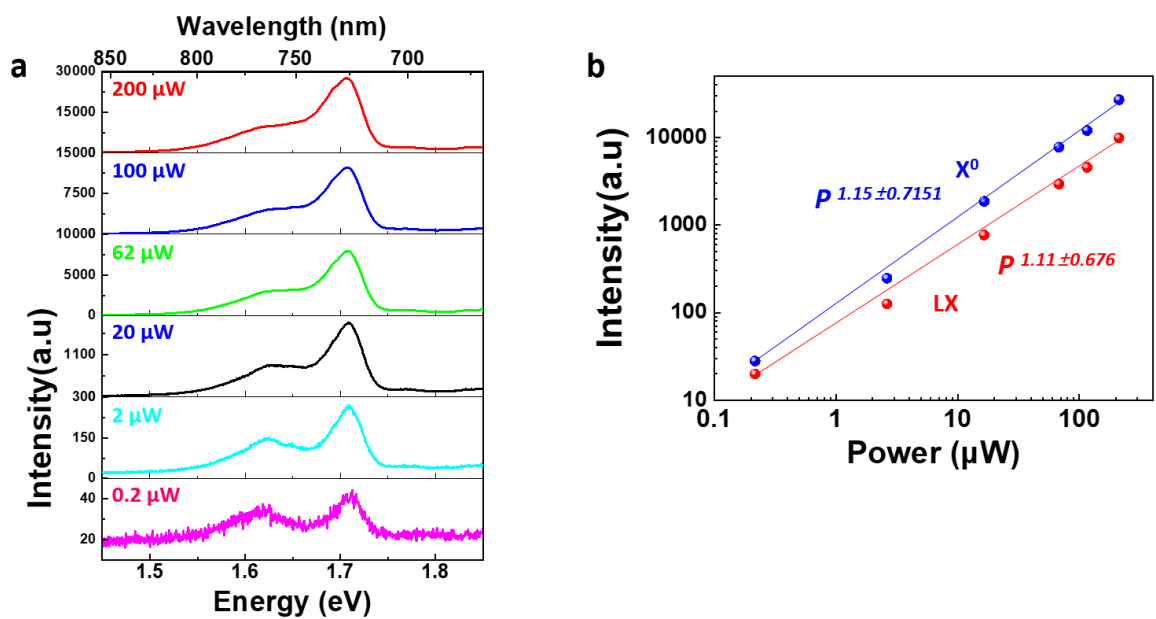

**Figure S13.** Laser power-dependent PL spectra of strained $WSe_2$ on $AlO_x$(5 nm)/Pt NPs/$SiO_2$ measured at 80 K. These data present the photoexcitation power-dependent PL spectra and a logarithmic plot of the PL intensity of $X^0$ and LX emissions measured at 80 K based on curve fitting the PL spectra. From the fit to the slope, the linear power-dependent factors are determined to be approximately 1.15 and 1.11 for $X^0$ and LX emissions, which excludes any possibility of emission by multi-particle complexes[1-4].

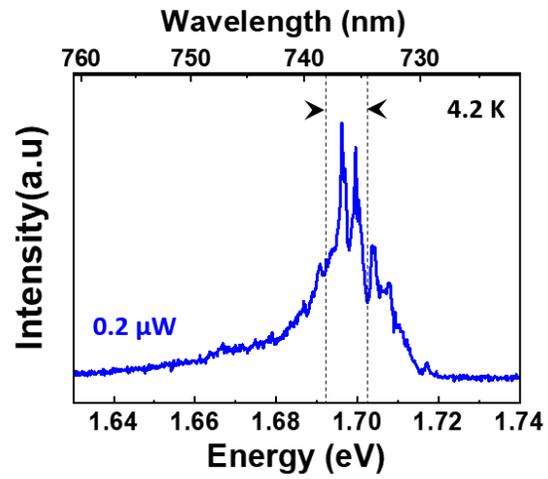

**Figure S14.** PL spectra of strained WSe$_2$ on AlO$_x$(5 nm)/Pt NPs/SiO$_2$ measured with pulsed excitation laser (637 nm, 200 nW, 1 MHz) at 4.2K. The peaks have sharp linewidths of (675 ± 118 μeV).

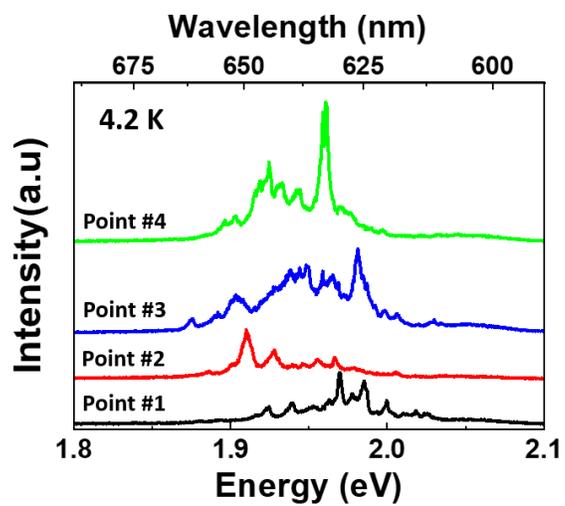

**Figure S15.** PL spectra of strained $WS_2$ on $AlO_x$(5 nm)/Pt NPs/$SiO_2$ measured with pulsed excitation laser (470 nm, 5 µW, 100 MHz) at 4.2K. The peaks have sharp linewidths of 728 ± 129 µeV.